# Calculation of Electron Swarm Parameters in Ar/H$_2$ and Kr/H$_2$ Mixtures

Mohamed Mostefaoui*, Djilali Benyoucef, Rachid Taleb, Abdelatif Gadoum
Laboratoire Génie Electrique et Energies Renouvelables, Université Hassiba Benbouali de Chlef,
Chlef, Algeria
*ostefm@gmail.com

*Abstract*—To develop the modeling of electrical discharges in a gaseous medium, it must to know a numbers of basic data such as: the electron collision cross sections for particle modeling or the electron swarm parameters in the case of the fluid modeling. The knowledge of all the electron swarm parameters as a function of the reduced electric field is indispensable for the fluid models based on the first and the second moment of the electron energy distribution function. The objective of this work is to calculate the mobility, the characteristic energy, and the ionization coefficient in the case of the mixtures Ar/H2 and Kr/H2. The results obtained will be compared with measurements already made by different authors.

## I. INTRODUCTION

The Boltzmann equation is the basic kinetic equation for the description of noble gases. This equation describes the evolution of the gas in the microscopic scale, where the average free path is of the same order of magnitude as a macroscopic length characteristic of the physical domain considered. When the average free path becomes small in a comparison to this characteristic length, a macroscopic description of the particle system is better adapted. In this case, the modeling gas state is then done by using the local average quantities which characterize it, such as mass density, temperature (or energy) and average velocity, characteristic energy, and the ionization coefficient. Otherwise, the adopted model is the particle model [1-6]. The objective of this work is to calculate the mobility, the characteristic energy, and the ionization coefficient in the case of the mixtures Ar/H$_2$ and Kr/H$_2$. The results obtained will be compared with measurements already made by different authors.

## II. RESOLUTION OF THE BOLTZMANN EQUATION AND DERIVATION OF THE ELECTRON SWARM PARAMETERS

### A. Boltzmann Equation

The Boltzmann equation for a set of electrons in an ionized gas under the electric field effect is given by the following equation:

$$\frac{\partial f_e}{\partial t} = \mathbf{v}.\nabla f_e - \frac{e}{m_e}\mathbf{E}.\nabla_v f_e = C[f_e] \quad (1)$$

where $f_e$ is the electron distribution function in the six-dimensional phase space (3 for ordinary space and 3 for velocity components), $\mathbf{v}$ is electron velocity, $e$ is the elementary charge, $m_e$ is the electron mass, $E$ is the electric field, $\nabla_v$ the operator of the velocity gradient, and $C[f_e]$ due to the electrons production and loss by electrons collisions with heavy species. To solve the Boltzmann Equation (BE), we need to make radical simplifications. In the beginning, we limit ourselves to the case where the electric field and the collision probabilities are all spatially uniform, at least on the scale of the average free path. The electron distribution function $f_e$ is then symmetrical in the velocity space around the direction of the electric field. In the position space, the distribution function $f_e$ can be varied only in the direction of the electric field. In this condition, and by using the spherical coordinates in the velocity space, equation (1) can be written as follow:

$$\frac{\partial f_e}{\partial t} + \upsilon \cos\theta \frac{\partial f_e}{\partial z} - \frac{e}{m_e}E\left(\cos\theta \frac{\partial f_e}{\partial \upsilon} + \frac{\sin^2\theta}{\upsilon}\frac{\partial f_e}{\partial \cos\theta}\right) = C[f_e] \quad (2)$$

where $\upsilon$ is the velocity amplitude, $\theta$ is the angle between the velocity and the direction of the electric field and $z$ is the position along this direction. The electron distribution function $f_e$ in equation (2) depends on four coordinates: $\upsilon$, $\theta$, $t$ and $z$. The next sections describe how we deal with this. We simplify the dependence $\theta$ by the classical two-term approximation method. To simplify the temporal dependence, we consider only the cases of stable state where the electric field and the electron distribution function are stationary. We then describe the term collision, by putting all the elements together in one equation.

### B. two-term approximation method

A common approach to solve equation (2) is to develop $f_e$ in Legendre polynomial terms of $\cos\theta$ (expansion of spherical harmonics), and then to construct from equation (2) a set of equations for expansion coefficients. To obtain high precision results, six or more expansion terms are needed [7], but in many cases, a two-term approximation already gives useful results. This two-term approximation is often used (for example by solvers BOLSIG and ELENDIF) and has been widely discussed in the literature [8, 9].

Although it is known that, the approximation fails for high E/N values when most collisions are inelastic and $f_e$ becomes strongly anisotropic [10], errors in transport coefficients and rate coefficients are acceptable for fluid discharge modeling in the usual range of discharge conditions.

Note that when the two-term approximation fails, some other, intrinsic approximations of fluid models also fail. Nom by using the two-term approximation, the distribution function $f_e$ can be written as flow:

$$f_e(\upsilon, \cos\theta, z, t) = f_{e,0}(\upsilon, z, t) + f_{e,1}(\upsilon, z, t)\cos\theta \quad (3)$$

where $f_{e,0}$ is the isotropic part of the electron distribution function $f_e$ and $f_{e,1}$ is the anisotropic perturbation. Note that $\theta$ is defined relative to the direction of the electric field, so $f_{e,1}$ is negative; this differs from some other texts where $\theta$ is defined with respect to the drift velocity of the electrons and $f_{e,1}$ is positive, the electron distribution function $f_e$ is normalized as:

$$\iiint_\upsilon f_e \, d^3\upsilon = 4\pi \int_0^\infty f_{e,0} \upsilon^2 d\upsilon = n_e(z,t) \quad (4)$$

where $n_e$ is the electrons density number. The equations for $f_{e,0}$ and $f_{e,1}$ are found in equation (2) by substituting equation (3), and multiplying by the respective Legendre polynomials (1 and $\cos\theta$) and integrating on $\cos\theta$, we obtained the following equations:

(5)

where $\gamma = (2e/m)^{1/2}$ is a constant and $\varepsilon = (\upsilon/\gamma)^2$ is the electron energy in electron-volts, and $N$ is the neutral density. The right side of the first equation (5) represents the change of $f_{e,0}$ due to collisions. The right side of the second equation (5) contains the total momentum transfer cross-section $\sigma_m$ consisting of the contributions of all possible processes $k$ with gas particles:

$$\sigma_m = \sum_k p_k \sigma_k \quad (6)$$

where $p_k$ is the molar fraction of the target species of the collision process; realize that the gas can be a mixture of different species, including excited states. For elastic collisions, $\sigma_k$ is the moment transfer cross section [5], taking into account the possible anisotropy of elastic scattering. For inelastic collisions, $\sigma_k$ is the total cross section, assuming that all momentum is lost in the collision, and after the collision, the remaining electronic velocity collision is scattered isotropically.

C. *Electron density increasing*

The Equation (5) is simplified by making assumptions about the temporal and spatial dependence of $f_{e,0}$ and $f_{e,1}$. In general, $f_e$ cannot be constant in time and space because some collision processes (ionization, attachment) do not retain the total number of electrons. In the references [8, 11-13] there is a simple technique proposed to roughly describe the effects of net electron production in swarms. By this technique, the distribution function $f_e$ can be separated from its dependence on time and space.

$$f_{e,0,1}(\varepsilon, z, t) = \frac{1}{2\pi\gamma^3} F_{e,0,1}(\varepsilon) n_e(z,t) \quad (7)$$

where the electron distribution energy $F_{e,0,1}$ is constant in the time and the space, and its normalization is given by the following expression:

$$\int_0^\infty \sqrt{\varepsilon} F_{e,0}(\varepsilon) d\varepsilon = 1 \quad (8)$$

The temporal or spatial dependence of the electron density is now related to the net electron production rate. For this, we consider two simple cases corresponding to specific swarm experiments. Exponential temporal growth without spatial dependence, this case corresponds to the experiences of Pulsed Townsend [14]. The temporal growth rate of the density of the number of electrons is equal to the net production frequency ($v_i$) given by the following equation:

$$\frac{1}{n_e}\frac{\partial n_e}{\partial t} = \overline{v}_i = N\gamma \int_0^\infty \left( \sum_{k_i} p_{k_i}\sigma_{k_i} - \sum_{k_a} p_{k_a}\sigma_{k_a} \right) \varepsilon F_{e,0} d\varepsilon \quad (9)$$

where the sum is on the processes of ionization and attachment, we recall that $p_k$ is the molar fraction of the target species of the collision process $k$ From equation (5), the anisotropic part of the electron energy distribution function becomes:

$$F_1 = \frac{E}{N}\frac{1}{\tilde{\sigma}_m}\frac{\partial F_{e,0}}{\partial \varepsilon}$$
$$\tilde{\sigma}_m = \sigma_m + \frac{\overline{v}_i}{N\gamma\sqrt{\varepsilon}} \quad (10)$$

By substituting $F_{e,1}$ (eq.10) in isotropic part of equation (5), we find:

$$-\frac{\gamma}{3}\frac{\partial}{\partial \varepsilon}\left[\left(\frac{E}{N}\right)^2 \frac{\varepsilon}{\tilde{\sigma}_m}\frac{\partial F_{e,0}}{\partial \varepsilon}\right] = \tilde{C}_0 + \tilde{R} \quad (11)$$

where the collision term is given by the following expression:

$$\tilde{C}_0 = 2\pi\gamma^3 \sqrt{\varepsilon}\frac{C_0}{Nn}$$
$$\tilde{R} = -\frac{\overline{v}_i}{N}\sqrt{\varepsilon}F_0 \quad (12)$$

In the reference [12] there is an interpretation of this term as the energy needed to heat secondary electrons to the average electron energy. The Exponential spatial growth without temporal dependence is consistent with the Steady State Townsend experiments [14]. As electrons drift against the electric field, their flux and density increase exponentially with a constant spatial growth rate $\alpha$ (Townsend coefficient), which is related to net electron production by:

$$\alpha = -\frac{1}{n_e}\frac{\partial n_e}{\partial z} = -\frac{\bar{\nu}_i}{\omega} \quad (13)$$

where the average velocity $\omega$ is determined by $F_{e,1}$, it is constant in space with a negative value. By substituting $\alpha$ (eq.13) in anisotropic part of equation (5), we find:

$$F_{e,1} = -\frac{1}{\sigma_m}\left(\frac{E}{N}\frac{\partial F_{e,0}}{\partial \varepsilon} + \frac{\alpha}{N}F_{e,0}\right) \quad (14)$$

Finally, the isotropic part of the electron energy distribution function (eq.5) can be written as follows:

$$-\frac{\gamma}{3}\frac{\partial}{\partial \varepsilon}\left[\left(\frac{E}{N}\right)^2 \frac{\varepsilon}{\sigma_m}\frac{\partial F_{e,0}}{\partial \varepsilon}\right] = 2\pi\gamma^3\sqrt{\varepsilon}\frac{C_0}{Nn} + \frac{\alpha}{N}\frac{\gamma}{3}\left[\frac{\varepsilon}{\sigma_m}\left(2\frac{E}{N}\frac{\partial F_{e,0}}{\partial \varepsilon}\frac{\alpha}{N}F_0\right) + \frac{E}{N}F_{e,0}\frac{\partial}{\partial \varepsilon}\left(\frac{\varepsilon}{\sigma_m}\right)\right] \quad (15)$$

The first Townsend coefficient $\alpha$ can be found by combining equation (13) and equation (14):

$$\omega = -\frac{1}{3}\gamma\int_0^\infty F_1\varepsilon\,d\varepsilon = -\mu E + \alpha D = -\frac{\bar{\nu}_i}{\alpha} \quad (16)$$

This leads to:

$$\alpha = \frac{1}{2D}\left(\mu E - \sqrt{(\mu E)^2 - 4D\bar{\nu}_i}\right) \quad (17)$$

where $\mu$ et $D$ are respectively, the mobility and the diffusion coefficient, then, the reduced mobility $\mu N$ and characteristic energy (reduced diffusion $DN$) can be calculated by the following expressions:

$$\mu N = -\frac{\gamma}{3}\int_0^\infty \frac{\varepsilon}{\tilde{\sigma}_m}\frac{\partial F_0}{\partial \varepsilon}d\varepsilon \quad (18)$$

$$DN = \frac{\gamma}{3}\int_0^\infty \frac{\varepsilon}{\tilde{\sigma}_m}F_0\,d\varepsilon \quad (19)$$

III. RESULTS AND DISCUSSION

The cross sections of Argon, Krypton, and Hydrogen used in this work are those of Phelps [15], the results are obtained for a quasi-stationary electric field. The results obtained are compared with the measurements already exists in the literature [16-44].

A. Ar/H₂ Mixture

The following figure shows the mobility, the characteristic energy, and the ionization coefficient in the Ar/H₂ mixture as a function of the reduced electric field $E/N$ under the following conditions :

1: Ar 100%/H₂ 0%
2: Ar 75%/H₂ 25%
3: Ar 50%/H₂ 50%
4: Ar 25%/H₂ 75%
5: Ar 0%/H₂ 100%

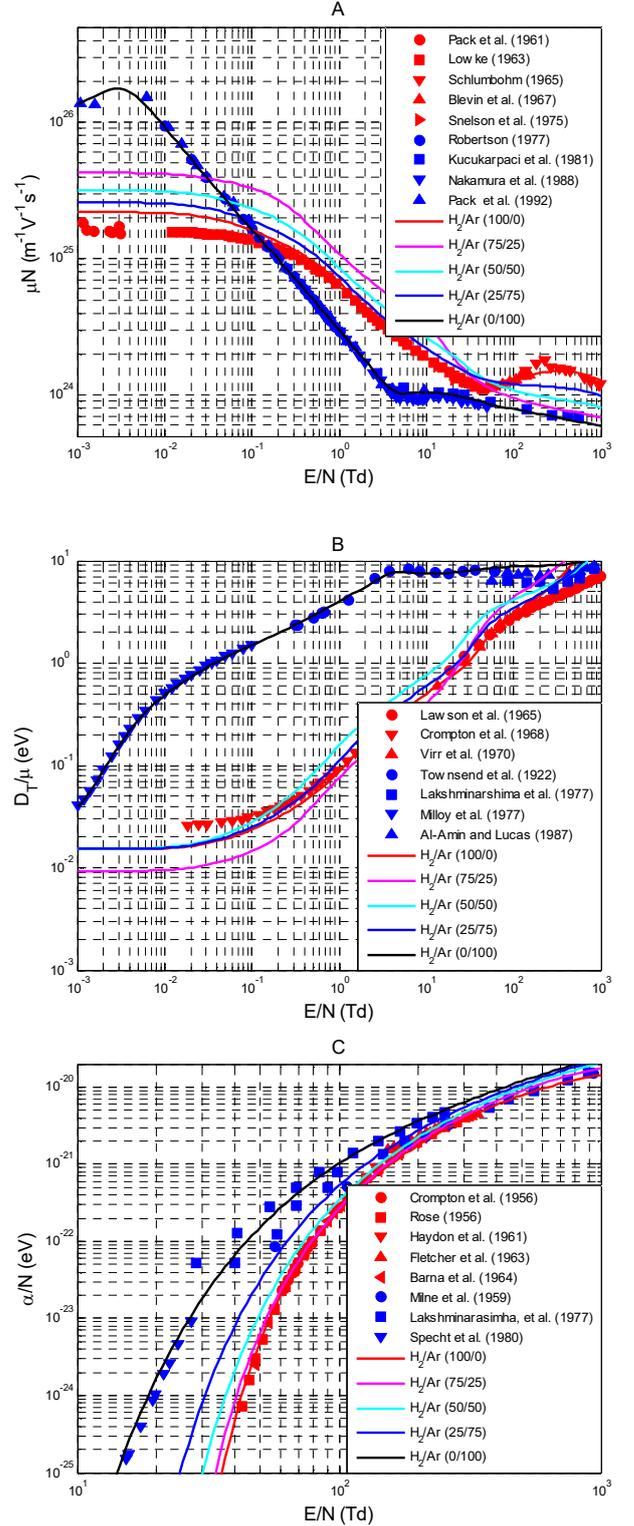

Fig.1. Comparison between calculated swarm parameters and those measured in the mixture (Ar/H₂); (A) The characteristic energy of electrons, (B): Electron mobility, (C) Ionization coefficient

*B. Kr/H₂ Mixture*

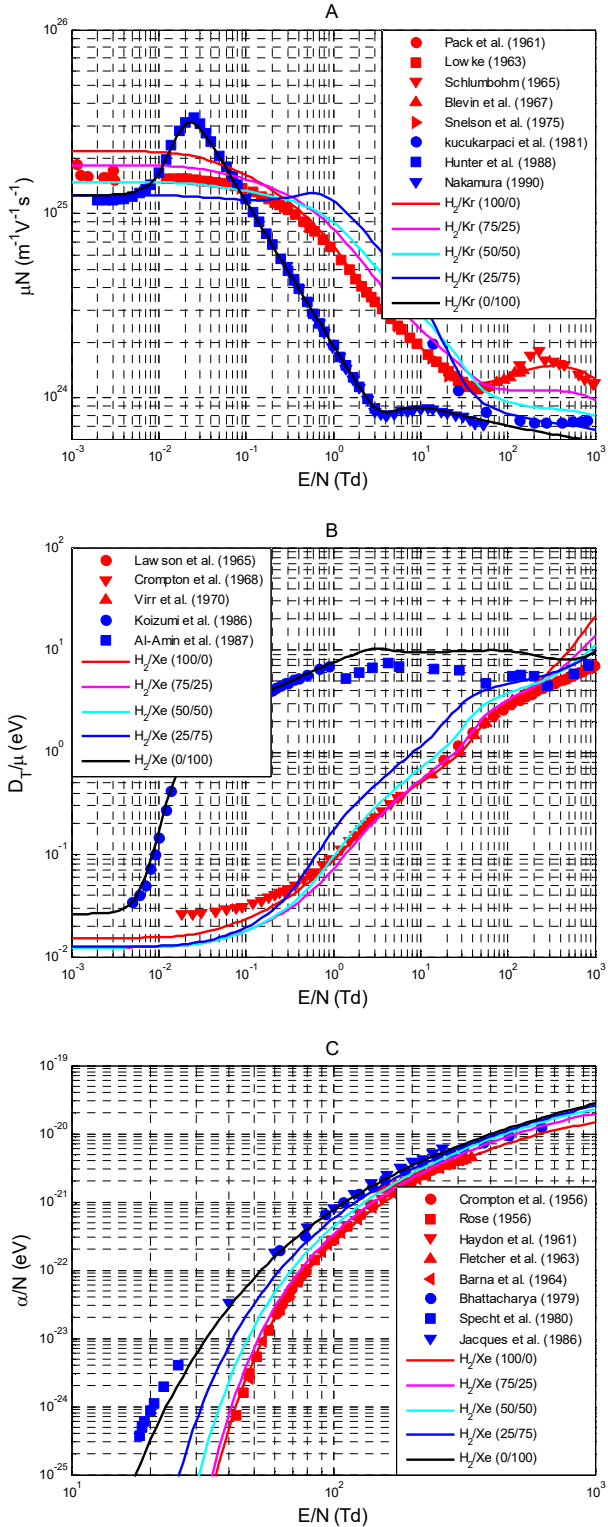

Fig.2. Comparison between calculated swarm parameters and those measured in the mixture (Kr/H₂); (A) The characteristic energy of electrons, (B): Electron mobility, (C) Ionization coefficient

In Figure 2, we show the mobility, the characteristic energy, and the ionization coefficient in the Kr/H₂ mixture as a function of the reduced *E/N* field under the following conditions:

1: Kr 100%/$H_2$ 0%
2: Kr 75%/$H_2$ 25%
3: Kr 50%/$H_2$ 50%
4: Kr 25%/$H_2$ 75%
5: Kr 0%/$H_2$ 100%

The results obtained for the pure gases of Argon, Krypton, and Hydrogen are in very good agreement with the measurements data, where for low electric field the electron mobility, the characteristic energy, and also the ionization coefficient decrease by increasing the ratio of hydrogen, this is can be explained by the lower excited state of hydrogen (rotational states and vibrational states). These last, absorbs the electrical energy gained by the electrons movement in the electric field, this decrease is due as a result of inelastic collisions with the hydrogen molecule.

IV. CONCLUSION

In this work we have shown the transport coefficients in the Argon/Hydrogen and Krypton/Hydrogen mixtures. The results obtained for the pure gases of Argon, Krypton, and Hydrogen are in very good agreement with the measurements. In general, we note that the ionization coefficient increases with the increase of the percentage of the noble gases (Ar, Kr). For this reason, these gases are used as additive with the any gas to increase the density of the electrons in the electric discharges.